\documentclass{article}

\usepackage{amssymb}

\newcommand{\Pc}{\mathcal{P}}

\newcommand{\Wc}{\mathcal{W}}

\newcommand{\id}{\mathbf{I}}

\newcommand{\Tr}{\mbox{Tr}}

\newtheorem{theor}{Theorem}
\newtheorem{prop}{Proposition}
\newtheorem{lemma}{Lemma}
\newtheorem{definition}{Definition}

\newenvironment{proof}{
\noindent
{\bf Proof.}
}{
\begin{flushright}$\blacksquare$\end{flushright}
}

\title{The Secrecy Capacity of the MIMO Wiretap Channel
}
\author{Fr\'ed\'erique Oggier and Babak Hassibi
\footnote{The authors are with 
Department of Electrical Engineering, 
California Institute of Technology,
 Pasadena 91125 CA, USA.
Email:{\tt\small \{frederique,hassibi\}@systems.caltech.edu}
This work was supported in part by  NSF grant CCR-0133818,
by Caltech's Lee Center for Advanced Networking and by a grant 
from the David and Lucille Packard Foundation.
}
}
\begin{document}
\maketitle

\begin{abstract}
We consider the MIMO wiretap channel, that is a MIMO broadcast channel
where the transmitter sends some confidential 
information to one user which is a legitimate receiver, while the
other user is an eavesdropper. Perfect secrecy is achieved when the
the transmitter and
the legitimate receiver can communicate at some positive rate, while 
insuring that the eavesdropper gets zero bits of information. In this
paper, we compute the perfect secrecy capacity of the multiple 
antenna MIMO broadcast channel, where the number of antennas is 
arbitrary for both the transmitter and the two receivers.
\end{abstract}

%
%

\section{Introduction}

Security in wireless communication is a critical issue, which has
recently attracted a lot of interest. By nature, wireless channels
offer a shared medium, particularly favorable to eavesdropping.  
Among the numerous points of view from which security has been 
investigated, we adopt here the one of {\em information theoretic} 
security. In this context, most of the works dealing with wireless 
communication are based on the seminal work of Wyner \cite{Wyner}, and
its model, {\em the wire-tap channel}. 

\subsection{Information theoretic confidentiality}

In a traditional confidentiality setting, a transmitter (Alice) wants to
send some secret message to a legitimate receiver (Bob), and
prevent the eavesdropper (Eve) to have knowledge of the message. 

From an information theoretic point of view, the communication channel
involved can be modeled as a broadcast channel, following the 
wire-tap channel model introduced by Wyner \cite{Wyner}: a transmitter 
broadcasts its message, say $w^k \in \Wc^k$, encoded into a codeword 
$x^n$, and the two receivers (the legitimate and the illegitimate) 
respectively receive $y^n$ and $z^n$, the output of their channel. 
The knowledge that the eavesdropper gets of $w^k$ from
its received signal $z^n$ is modeled by  
\[
I(z^n;w^k)=h(w^k)-h(w^k|z^n),
\]
since the mutual information measures the amount of information that 
$z^n$ contains about $w^k$. The notion of {\em perfect secrecy} captures
the idea that whatever are the resources available to the
eavesdropper, they will not allow him to get a single bit of information.  
Perfect secrecy thus requires
\[
I(z^n;w^k)=0 \iff h(w^k)=h(w^k|z^n).
\]
In other words, the amount of randomness is the same in $w^k$ or in 
$w^k|z^n$.

The decoder computes an estimate $\hat{w}^k$ of the transmitted
message $w^k$, and 
the probability $P_e$ of decoding erroneously is given by
\begin{equation}\label{eq:defPe}
P_e=Pr(w^k\neq\hat{w}^k ).
\end{equation}
The amount of ignorance that the eavesdropper has about a message 
$w^k$ is called the {\em equivocation rate}, and following the above
discussion, it is naturally defined as: 
\begin{definition}\label{def:Re}
The {\em equivocation rate} $R_e$ at the eavesdropper is
\[
R_e=\frac{1}{n}h(w^k|z^n),
\]
\end{definition}
with $0\leq R_e \leq h(w^k)/n$. 
Clearly, if $R_e$ is equal to the information rate $h(w^k)/n$, then 
$I(z^n|w^k)=0$, which yields perfect secrecy.

To perfect secrecy is associated a {\em perfect secrecy rate} $R_s$, which
is the amount of information that can be sent not only reliably but
also confidentially, with the help of a $(2^{nR_s},n)$ code. 
\begin{definition}\label{def:Rs}
A perfect secrecy rate $R_s$ is said to be {\em achievable} if for any 
$\epsilon >0$, there exists a sequence of $(2^{nR_s},n)$ codes such
that for any $n \geq n(\epsilon)$, we have
\begin{eqnarray}
P_e &\leq& \epsilon \label{eq:Pe} \\
R_s-\epsilon &\leq& R_e. \label{eq:Re}
\end{eqnarray}
\end{definition}
The first condition (\ref{eq:Pe}) is the standard
definition of achievable rate as far as reliability is concerned.
The second condition (\ref{eq:Re}) guarantees secrecy, up to the
equivocation rate, which we will require to be $h(w^k)/n$ to have
perfect secrecy.
The {\em secrecy capacity} is defined similarly to the standard 
capacity:
\begin{definition}\label{def:Cs}
The secrecy capacity $C_s$ is the maximum achievable perfect secrecy
rate. 
\end{definition}

\subsection{Previous work}

In his seminal work \cite{Wyner}, Wyner showed for discrete memoryless 
channels that the perfect secrecy capacity is actually the difference of 
the capacity of the two users. To prove this result, he worked under
the assumption that the channel of the eavesdropper is a degraded
version of the channel of the legitimate receiver. 
This result has been generalized to Gaussian channels by Leung et 
al. \cite{Hellman}, under the same assumption.
   
The wire-tap channel has been adopted as a model for numerous works 
on information theoretic security, and in particular for those on
fading channels, both for point-to-point and multi-user systems. We
mainly review the prior work for point-to-point. 
In \cite{elgamal}, Gopala et al. have shown that the
secrecy capacity is also the difference of the two capacities in the
case of a single antenna fading channel, under the assumption of
asymptotically long coherence intervals, when the transmitter either
knows both channels or only the legitimate channel. When only the
legitimate channel is known, an optimal power allocation is given,
using a variable rate transmission scheme. 
In \cite{barros}, Barros et al. have characterized information theoretic
security in terms of outage probability. In the case when the
transmitter does not know the eavesdropper channel, they define the
probability of transmitting at a secrecy rate $R_S$ bigger than the
secrecy capacity $C_S$ (i.e. the outage probability) as the
probability that the information theoretic security is
compromised. They compute this probability, and also show that the
probability that the secrecy capacity $C_S$ is positive can actually
be positive even if the average SNR of the legitimate channel is
weaker than the one of the eavesdropper. They extend their work in
\cite{bloch}, where they also consider the cases when Alice has either
imperfect or perfect knowledge of the eavesdropper channel. 
Independently, Liang et al. \cite{Poor} and Li et al. \cite{Li2} have
computed the secrecy capacity for the parallel wiretap channel with
independent subchannels, and derived optimal source power allocation. 
The secrecy capacity of the wiretap channel with single antenna fading
channel follows. Finally, the results of \cite{Poor} are extended in 
\cite{liang}, where a fading broadcast channel with confidential
messages is considered, with common information for two receivers, and
confidential information intended for only one receiver. The secrecy
capacity is computed for the parallel broadcast channel with both
independent and degraded subchannels. 

In this work, we are interested in the perfect secrecy capacity of  
multiple antenna channels. A first study of the problem has been
proposed by Hero \cite{hero}. In a different context than the wire-tap
channel, he introduced the so-called constraints 
of low probability of detection, and low probability of intercept,
considering the scenario where the transmitter and the receiver are
both informed about their channel while the eavesdropper is uniformed 
about his. In \cite{Parada}, the SIMO wiretap channel has been
considered. Several results on the secrecy in MIMO communication have 
been provided very recently. In \cite{Li}, the secrecy capacity is
computed for the MISO case. Furthermore, a lower bound is computed in
the MIMO case. This lower bound, that is the achievability, is shown to be
the expected result, namely, the difference of the two channel
capacities, like in the previous cases. 
Finally, the secrecy capacity for the MISO case has been proven independently
by Khisti et al. \cite{Khisti}, where furthermore an upper bound is
given for the MIMO case, in a regime asymptotic in SNR.

The contribution of this paper is to compute the perfect secrecy
capacity of the multiple antenna wire-tap channel, for any number of
transmit/receive antennas, as well as for any SNR regime. One of the 
difficulties in studying the MIMO wire-tap channel is that the
broadcast MIMO channel is not degraded, an assumption which is crucial
in the proof of the converse in the original paper by Wyner (as well
as in the proofs presented in \cite{Hellman,elgamal,barros,Poor}). 
In order to compute the secrecy capacity, we provide a proof technique 
for the converse, which is different than the original one, and 
allows us to deal with channels that are not degraded. 
Note that our result shows that the inner bound by Li et al. \cite{Li} 
is tight, and this is proved by the computation of an upper bound 
that actually matches the lower bound.

\subsection{The MIMO wiretap channel}

We consider the MIMO wiretap channel, that is, a broadcast channel
where the transmitter is equipped with $n$ transmit antennas, while
the legitimate receiver and an eavesdropper have respectively $n_M$ and 
$n_E$ receive antennas. Thus, our model is described by the following
broadcast channel 
\begin{eqnarray*}
Y &=&H_MX+V_M \\
Z &=&H_EX+V_E 
\end{eqnarray*}
where  $Y,V_M$ and $Z,V_E$  are respectively $n_M\times 1$ and 
$n_E \times 1$ vectors. The notation 
that we will use throughout the paper is that the subscript $M$ refers
to the main channel (the one of the legitimate receiver), while the 
subscript $E$ refers to the eavesdropper channel. We will denote by 
$\id_n$ the $n\times n$ identity matrix, and by ${\bf 0}_n$ the 
$n\times n$ all zero matrix. We may omit the subscript if the
dimension is obvious.

We make the following assumptions:
\begin{itemize}
\item
$X$ is the $n\times 1$ transmitted signal, with covariance matrix 
$K_X \succeq {\bf 0}_n$ satisfying the power constraint 
\[
\Tr(K_X)=P.
\]
The power constraint holds for the whole paper, and we may sometimes
omit to repeat it explicitly.
\item
$H_M$ and $H_E$ are respectively $n_M\times n$ and $n_E\times n$ 
fixed channel matrices such that
\[
H_M^*H_M \succ {\bf 0}_n,~H_E^*H_E \succ {\bf 0}_n.
\] 
They are assumed to be known at the transmitter.
\item
$V_M,V_E$ are independent circularly symmetric complex Gaussian
 vectors with identity covariance $K_M=\id_{n_M}$, $K_E=\id_{n_E}$ 
and independent of the transmitted signal $X$. 
\end{itemize}

\begin{theor}
The secrecy capacity of the MIMO wiretap channel is given by
\[
C_S=\max_{K_X\succeq {\bf 0}}
\log\det(\id+H_MK_XH_M^*)-\log\det(\id+H_EK_XH_E^*)
\]
\end{theor}
where $\Tr(K_X)=P$. 
The paper contains the proof of the above theorem: in Section 
\ref{sec:achiev}, we prove an achievability 
result which characterizes the optimal matrices $\tilde{K_X}$, while 
Section \ref{sec:conv} contains the main results, namely the proof of
the converse. 

%
%
\section{On the Achievability}\label{sec:achiev}

In this section, we state the achievability part of the secrecy
capacity, and further prove that in the non-degraded case, the 
achievability is maximized by $n\times n$ matrices $K_X$ which are 
low rank, that is of any rank $r < n$.
\begin{prop}
The perfect secrecy rate 
\[
R_s=\max_{K_X \succeq {\bf 0},\Tr(K_X)=P}
\log\det(\id+H_MK_XH_M^*)-\log\det(\id+H_EK_XH_E^*)
\]
is achievable.
\end{prop}
This has already been proved \cite{Li}. In fact, the
interpretation is obvious. When $K_X$ is chosen, the difference 
between the resulting mutual informations to the legitimate 
user and eavesdropper can be secretly transmitted.

\begin{prop}\label{prop:lowrankach} 
Let $\tilde{K}_X$ be an optimal solution to the optimization problem
\begin{eqnarray*}
\max K_X & \log \det(\id +H_MK_XH_M^*)-\log \det (\id+H_EK_XH_E^*)\\
\mbox{s.t.} &  K_X \succeq {\bf 0},~\Tr(K_X)=P,  
\end{eqnarray*}
where $H_E^*H_E- H_M^*H_M$ is either indefinite or semidefinite. 
Then $\tilde{K}_X$ is a low rank matrix.
\end{prop}
\begin{proof}
In order to show that the optimal $\tilde{K}_X$ is low rank, 
we define a Lagrangian which includes the power constraint, 
and show that this yields no solution. From there, we can 
conclude that the optimal solution is on the boundary of the 
cone of positive semi-definite matrices, namely matrices of rank 
$r<n$.

We thus define the following Lagrangian:
\[
\log \det(\id_{n_M} +H_MK_XH_M^*)-\log \det (\id_{n_E}+H_EK_XH_E^*)
-\lambda\Tr(K_X),
\]
and look for its stationary points, that is for the solution 
of the following equation:
\begin{equation} \label{eq:devzero}
\begin{array}{l}
\nabla_{K_X}(\log \det(\id +H_MK_XH_M^*)
-\log \det (\id+H_EK_XH_E^*)-\lambda\Tr(K_X))=0 \\
\iff ((H_M^*H_M)^{-1}+K_X)^{-1}=((H_E^*H_E)^{-1}+K_X)^{-1}+\lambda \id_n.
\end{array}
\end{equation}
By pre-multiplying the above equation by $(K_X+(H_M^*H_M)^{-1})$ and
post-multiplying it by $(K_X+(H_E^*H_E)^{-1})$, we get 
\[
(H_E^*H_E)^{-1}+K_X = 
(H_M^*H_M)^{-1}+K_X+\lambda((H_M^*H_M)^{-1}+K_X)((H_E^*H_E)^{-1}+K_X),
\]
or equivalently
\begin{equation}\label{eq:pdm}
((H_E^*H_E)^{-1}-(H_M^*H_M)^{-1})\frac{1}{\lambda}=
((H_M^*H_M)^{-1}+K_X)((H_E^*H_E)^{-1}+K_X).
\end{equation}
Now, we have by assumption that $H_M^*H_M\succ {\bf 0}_n$ and 
$H_E^*H_E\succ {\bf 0}_n$. 
If furthermore $K_X\succ {\bf 0}$, then all the eigenvalues of 
$((H_M^*H_M)^{-1}+K_X)((H_E^*H_E)^{-1}+K_X)$ are strictly positive 
(see Lemma \ref{lem:spd}, in Appendix). 
This implies that (\ref{eq:pdm}) can have a 
solution if and only if the Hermitian matrix 
$((H_E^*H_E)^{-1}-(H_M^*H_M)^{-1})\frac{1}{\lambda}$ is positive definite. 
This means that either $H_M^*H_M\succ H_E^*H_E$ and 
$\lambda >0$, or $H_M^*H_M \prec H_E^*H_E$ and $\lambda <0$.
This gives a contradiction if $H_M^*H_M-H_E^*H_E$ is either indefinite
or semidefinite, implying that $\tilde{K_X}$ has to be low rank. 
\end{proof}

%
%
\section{Proof of the Converse}\label{sec:conv}

The goal of this section is to prove the converse, namely
\begin{theor}
For any sequence of $(2^{nR_s},n)$ codes with probability of 
error $P_e \leq\epsilon$ and equivocation rate 
$R_s-\epsilon \leq R_e$ for any $n\geq n(\epsilon)$,  
$\epsilon>0$, then the secrecy rate $R_s$  satisfies
\[
R_s \leq \max_{K_X\succeq {\bf 0},\Tr(K_X)=P}\log\det(\id+H_MK_XH_M^*)-
\log\det(\id+H_EK_XH_E^*).
\]
\end{theor}
The proof is done in three main steps, that we briefly 
sketch before entering into the details. 

First (subsection \ref{sec:step1}),  
we have, similarly to \cite{Hellman,elgamal} that 
\[
R_s-\epsilon \leq \frac{1}{n}[I(X^n;Y^n|Z^n)+\delta],~\epsilon, \delta >0.
\]
Thus, all the work consists of finding an upper bound on
$I(X;Y|Z)$. We will prove the following upper bound:
\[
I(X;Y|Z)\leq\max_{K_X\succeq {\bf 0}} \tilde{I}(X;Y|Z),
\]
where
\[
\begin{array}{c}
\tilde{I}(X;Y|Z)=
\log\det\left(
\id_n+
(H_M^*,~H_E^*)
\left(
\begin{array}{cc}
\id_{n_M} &  A \\
A^* & \id_{n_E}  
\end{array}
\right)^{-1}
\left(
\begin{array}{c}
H_M \\
H_E
\end{array}
\right)
K_X 
\right)\\
-\log\det(\id+H_EK_XH_E^*)
\end{array}
\]
and $A$ is an $n_M\times n_E$ matrix which denotes the correlation 
between $V_M$ and $V_E$.
At this point of the proof, the converse can be proved for the 
two ``simple'' cases when $H_M^*H_M \succ H_E^*H_E$ and 
$H_E^*H_E \succ H_M^*H_M$, 
which are the cases when the channel is degraded.

In general, $V_M$ and $V_E$ are independent. However, since the
secrecy capacity does not depend on $A$, we can assume that 
$\tilde{I}(X;Y|Z)$ is a function of both $A$ and $K_X$ for the
purposes of tightening our upper bound . We show 
(subsection \ref{sec:step2}) that $\tilde{I}(X;Y|Z)$ 
is actually concave in $K_X$ 
and convex in $A$. As a result, we obtain a new upper bound
\[
I(X;Y|Z)\leq \max_{K_X\succeq {\bf 0}} \tilde{I}(X;Y|Z),
\]
for all $A$ such that $\id-AA^* \succ {\bf 0}_{n_E}$, thus
\begin{eqnarray*}
I(X;Y|Z)&\leq & \min_A \max_{K_X\succeq {\bf 0}} \tilde{I}(X;Y|Z) \\
        &=    & \max_{K_X\succeq {\bf 0}}\min_A \tilde{I}(X;Y|Z).
\end{eqnarray*}
Furthermore, we jointly optimize $\tilde{I}(X:Y|Z)$ over 
$K_X$ and $A$, and compute the optimal $\tilde{A}$ in closed form 
expression, while showing that the optimal $\tilde{K}_X$ 
is on the boundary of its domain, namely, $\tilde{K}_X$ 
is low rank.

We conclude the proof (subsection \ref{sec:step3}) 
by showing that the converse matches 
the achievability.

\subsection{Bound on $I(X;Y|Z)$ and result for the degraded case}
\label{sec:step1}

We start by recalling a standard result, which has already 
been proved in \cite{Hellman,elgamal}.
\begin{lemma}
Given any sequence of $(2^{nR_s},n)$ codes with 
$P_e \leq\epsilon$ and $R_s-\epsilon \leq R_e$ for any 
$n\geq n(\epsilon)$, $\epsilon>0$, the secrecy rate $R_s$ 
can be upper bounded as follows:
\[
R_s-\epsilon \leq \frac{1}{n}[I((X^n,Y^n|Z^n)+\delta],
\]
for $\epsilon,\delta>0$.
\end{lemma}

We thus focus now on finding an upper bound on $I(X;Y|Z)$. 
We provide two approaches:
\begin{enumerate}
\item An upper bound is given by assuming that the legitimate 
receiver knows both his channel and the one of the eavesdropper. 
\item The same upper bound can also be obtained as follows. 
Clearly, $I(X;Y|Z)$ is upper bounded by taking the maximum 
over all input distributions $\Pc(X)$:
\[
I(X;Y|Z)\leq \max_{\Pc(X)}I(X;Y|Z) = 
\max_{K_X\succeq {\bf 0}}\tilde{I}(X;Y|Z), 
\]
where $\tilde{I}(X;Y|Z)$ denotes the value of $I(X;Y|Z)$ 
when $\Pc(X)$ is optimal.  We will prove that the optimal 
distribution is Gaussian.
\end{enumerate}

\begin{prop}\label{prop:firstbound}
We have the following upper bound:
\[
\begin{array}{c}
I(X;Y|Z)\leq
\max_{K_X\succeq {\bf 0}}
\log\det\left(
\id_n+
(H_M^*,~H_E^*)
\left(
\begin{array}{cc}
\id &  A \\
A^*   & \id
\end{array}
\right)^{-1}
\left(
\begin{array}{c}
H_M \\
H_E
\end{array}
\right)
K_X 
\right)\\
-\log\det(\id+H_EK_XH_E^*),
\end{array}
\] 
where $A$ denotes the correlation between $V_M$ and $V_E$ and
satisfies $\id-AA^* \succ {\bf 0}$.
\end{prop}
\begin{proof}
An upper bound on $I(X;Y|Z)$ is obtained by assuming that 
the legitimate receiver knows both its channel and the one of the
eavesdropper. In this case, the capacity of the link between the
transmitter and the legitimate receiver is that of a MIMO system,
namely 
\[
\max_{K_X}
\log\det\left(
\id_n+
(H_M^*,~H_E^*)
\left(
\begin{array}{cc}
\id_{n_M} &  A \\
A^*   &  \id_{n_E}
\end{array}
\right)^{-1}
\left(
\begin{array}{c}
H_M \\
H_E
\end{array}
\right)
K_X 
\right).
\]
Now the channel we consider is degraded, and an upper 
bound is thus the difference of the two capacities, which yields 
the result. 

We now provide the alternative proof.
Clearly
\[
I(X;Y|Z)\leq \max_{\Pc(X)} I(X;Y|Z),
\]
where $\Pc(X)$ denotes the input distribution.
Now note that
\begin{eqnarray*}
I(X;Y|Z) & = & h(Y|Z)-h(Y|X,Z) \\
         & = & h(Y|Z)-h(X,Y,Z)+h(X,Z) \\
         & = & h(Y|Z)-h(X)-h(Y,Z|X) +h(X)+h(Z|X)\\
         & = & h(Y|Z)-h(V_E,V_M)+h(V_E).
\end{eqnarray*}
Thus the optimization problem we have to solve is
\[
\max_{\Pc(X)} h(X+V_M,X+V_E)-h(X+V_E).
\]
Using Proposition \ref{prop:maxgauss} (see Appendix), 
the optimal is given by choosing $X$ Gaussian.
Thus we have that
\begin{eqnarray*}
I(X;Y|Z)  & = & h(Y|Z)-h(V_E,V_M)+h(V_E)\\
          & = & h(Y,Z)-h(Z)-h(V_E,V_M)+h(V_E),
\end{eqnarray*}
which, when $X$ is Gaussian, is given by
\[
\log\det(K_{YZ})-\log\det(K_Z)-\log\det(K_{ME})+\log\det(K_E)
\]
where $K_{YZ}$, $K_Z$, $K_{ME}$ and $K_E=\id_{n_E}$ are covariance 
matrices, with
\[
K_{YZ}=
\left(
\begin{array}{cc}
H_MK_XH_M^*+\id_{n_M} & H_MK_XH_E^*+A \\
H_EK_XH_M^*+A^* & H_EK_XH_E^*+\id_{n_E}
\end{array}
\right),
\]
where $A$ denotes the correlation between $V_M$ and $V_E$,
and
\[
K_{ME}=
\left(
\begin{array}{cc}
\id_{n_M} & A \\
A^* & \id_{n_E}
\end{array}
\right).
\]
In order for $K_{ME}$ to be well defined, $A$ has to satisfy 
$\id-AA^*\succeq {\bf 0}$.

Thus we have
\[
\begin{array}{l}
\log\det\left(
\left(
\begin{array}{cc}
\id &  A \\
A^*   &  \id
\end{array}
\right)+

\left(
\begin{array}{c}
H_M \\
H_E
\end{array}
\right)
K_X(H_M^*,~H_E^*)
\right)
-\log\det(H_EK_XH_E^*+\id)\\
-\log\det(K_{ME})\\
=
\log\det\left(
\id+
\left(
\begin{array}{cc}
\id &  A \\
A^*   &  \id
\end{array}
\right)^{-1}
\left(
\begin{array}{c}
H_M \\
H_E
\end{array}
\right)
K_X (H_M^*,~H_E^*)
\right)
-\log\det(H_EK_XH_E^*+\id),
\end{array}
\]
where the second equality is well defined if we further require 
$\id-AA^*\succ {\bf 0}$.
The value of $I(X;Y|Z)$ when $X$ is Gaussian is denoted by 
$\tilde{I}(X;Y|Z)$:
\begin{equation}\label{eq:itilde1}
\begin{array}{c}
\tilde{I}(X;Y|Z)=\log\det\left(
\id+
(H_M^*,~H_E^*)
\left(
\begin{array}{cc}
\id &  A \\
A^*  & \id  
\end{array}
\right)^{-1}
\left(
\begin{array}{c}
H_M \\
H_E
\end{array}
\right)
K_X 
\right)\\
-\log\det(\id+H_EK_XH_E^*).
\end{array}
\end{equation}
\end{proof}

We can now conclude the proof of the converse for the ``simple'' cases
when $H_M^*H_M \succ H_E^*H_E$ or $H_E^*H_E \succ H_M^*H_M$.

\begin{prop}\label{lem:easy}
\begin{enumerate}
\item
If $H_M^*H_M \succ H_E^*H_E$, we have that
\[
I(X;Y|Z)\leq \max_{K_X\succeq {\bf 0}}\log\det(\id+H_MK_XH_M^*)-
\log\det(\id+H_EK_XH_E^*).
\]
\item
Vice versa, if $H_E^*H_E \succ H_M^*H_M$, we have that 
\[
I(X;Y|Z)= 0.
\]
\end{enumerate}
\end{prop}
\begin{proof}
Let us first compute another way of writing $\tilde{I}(X;Y|Z)$, as 
defined in (\ref{eq:itilde1}).
Note the following factorization:
\[
\left(
\begin{array}{cc}
\id_{n_M} & A \\
A^* & \id_{n_E}
\end{array}
\right)
=
\left(
\begin{array}{cc}
\id & A \\
{\bf 0} & \id
\end{array}
\right)
\left(
\begin{array}{cc}
\id- AA^* & {\bf 0} \\
{\bf 0} & \id
\end{array}
\right)
\left(
\begin{array}{cc}
\id & {\bf 0} \\
A^* & \id
\end{array}
\right)
\]
so that
\[
\left(
\begin{array}{cc}
\id & A \\
A^* & \id
\end{array}
\right)^{-1}
=
\left(
\begin{array}{cc}
\id & {\bf 0} \\
-A^* & \id
\end{array}
\right)
\left(
\begin{array}{cc}
(\id- AA^*)^{-1} & {\bf 0} \\
{\bf 0} & \id
\end{array}
\right)
\left(
\begin{array}{cc}
\id & -A \\
{\bf 0} & \id
\end{array}
\right)
\]
and we have that
\[
(H_M^*,~H_E^*)
\left(
\begin{array}{cc}
\id &  A \\
A^*  & \id  
\end{array}
\right)^{-1}
\left(
\begin{array}{c}
H_M \\
H_E
\end{array}
\right)
=
(H_M^*-H_E^*A^*)(\id-AA^*)^{-1}(H_M-AH_E)+H_E^*H_E.
\]
Thus 
\begin{eqnarray}
\tilde{I}(X;Y|Z)&=&
\log\det(
\id +((H_M^*-H_E^*A^*)(\id-AA^*)^{-1}(H_M-AH_E)+H_E^*H_E)K_X) \nonumber \\
&&
-\log\det(
\id+H_EK_XH_E^*
). \label{eq:itilde3}
\end{eqnarray}
\begin{enumerate}
\item
Since the secrecy capacity does not depend on the noise correlation
$A$, and that
\[
I(X;Y|Z) \leq \max_{K_X} \tilde{I}(X;Y|Z),
\]
for all $A$ such that $\id-AA^* \succ {\bf 0}$, 
we are free to take $A^*=H_E (H_M^*H_M)^{-1}H_M^*$.
Indeed, such $A$ does {\em not} depend on
a choice of $K_X$, and since $H_M^*H_M \succ H_E^*H_E$, $A$ satisfies
\[
\id-AA^*=\id-H_M(H_M^*H_M)^{-1}H_E^*H_E(H_M^*H_M)^{-1}H_M^*  \succ {\bf 0}.
\]
Finally, we are left to show that by replacing $A^*$
with $H_E (H_M^*H_M)^{-1}H_M^*$ in $\tilde{I}(X;Y|Z)$ 
indeed yields $\log\det(\id+H_MK_XH_M^*)-\log\det(\id+H_EK_XH_E^*)$.
Consider thus $\tilde{I}(X;Y|Z)$ as defined in (\ref{eq:itilde3}). 
It is enough to show that
\[
(H_M^*-H_E^*A^*)(\id-AA^*)^{-1}(H_M-AH_E)+H_E^*H_E=H_M^*H_M.
\]
We have that
\begin{eqnarray*}
&&(\id-AA^*)^{-1}\\
&=&(\id-H_M(H_M^*H_M)^{-1}H_E^*H_E(H_M^*H_M)^{-1}H_M^*)^{-1}\\
&=&\id+H_M(H_M^*H_M)^{-1}((H_E^*H_E)^{-1}-
                  (H_M^*H_M)^{-1})^{-1}(H_M^*H_M)^{-1}H_M^*
\end{eqnarray*}
using the matrix inversion lemma, so that
\begin{eqnarray*}
H_M^*(\id-AA^*)^{-1}H_M
&=&H_M^*H_M+((H_E^*H_E)^{-1}-(H_M^*H_M)^{-1})^{-1}\\
&=&H_M^*H_M+(\id-H_E^*H_E(H_M^*H_M)^{-1})^{-1}H_E^*H_E
\end{eqnarray*}
and finally
\[
\begin{array}{l}
(\id-H_E^*H_E(H_M^*H_M)^{-1})H_M^*(\id-AA^*)^{-1}
H_M(\id-(H_M^*H_M)^{-1}H_E^*H_E\\
=H_M^*H_M-H_E^*H_E.
\end{array}
\]
\item
Similarly if $H_E^*H_E \succ H_M^*H_M$, we are free to choose 
$A^*=H_E(H_E^*H_E)^{-1}H_M^*$, which 
satisfies 
\[
\id-AA^*=\id-H_M(H_E^*H_E)^{-1}H_M^*\succ {\bf 0}.
\]
Since $H_M^*-H_E^*A^*={\bf 0}$, we see from (\ref{eq:itilde3}) that 
\[
\tilde{I}(X;Y|Z)=0.
\]
\end{enumerate}
\end{proof}

The cases described in the lemma can be understood as a simple
generalization of the scalar case, since those are the degraded cases. 
When $H_M^*H_M \succ H_E^*H_E$, all links to 
the legitimate receiver are better, and the capacity is given by the
difference of the two capacities, while if $H_E^*H_E \succ H_M^*H_M$, 
then all links to the eavesdropper are better, and thus no positive
secrecy capacity can be achieved.

We are now left with the case when $H_M^*H_M-H_E^*H_E$ is indefinite, 
which is the non-degraded case, and thus the interesting case to 
understand.

\subsection{Minimization over $A$ and maximization over $K_X$}
\label{sec:step2}

We have shown in Proposition \ref{prop:firstbound} that
\[
I(X;Y|Z) \leq \max_{K_X\succeq {\bf 0}} \tilde{I}(X;Y,Z).
\]
Since this is true for all $A$ such that $\id-AA^*\succ {\bf 0}$, 
we further have that
\[
I(X;Y|Z) \leq \min_A\max_{K_X} \tilde{I}(X;Y,Z).
\]
To understand this double optimization, we start by analyzing the 
function $\tilde{I}(X;Y,Z)$.
\begin{prop}\label{prop:con}
The function $\tilde{I}(X;Y,Z)$ defined in (\ref{eq:itilde1})
is concave in $K_X$ and convex in $A$. 
Consequently,
\[
\min_A \max_{K_X} \tilde{I}(X;Y|Z)=\max_{K_X} \min_A \tilde{I}(X;Y|Z)
\]
where $K_X$ and $A$ respectively satisfy
\[
\Tr(K_X)=P,~K_X\succeq {\bf 0},~\id-AA^*\succ {\bf 0}.
\]
\end{prop}
\begin{proof}
Recall from (\ref{eq:itilde1}) that $\tilde{I}(X;Y|Z)$ is given by
\[
\log\det\left(
\id_n+
(H_M^*,~H_E^*)
\left(
\begin{array}{cc}
\id_{n_M} &  A \\
A^*   & \id_{n_E}
\end{array}
\right)^{-1}
\left(
\begin{array}{c}
H_M \\
H_E
\end{array}
\right)
K_X 
\right)
-\log\det(\id+H_EK_XH_E^*).
\]
\begin{enumerate}
\item
{\bf Convexity in $A$.}
Set
\[
C:=
\left(
\begin{array}{cc}
\id &  A \\
A^*   &  \id
\end{array}
\right),~
D:=
\left(
\begin{array}{c}
H_M \\
H_E
\end{array}
\right)K_X
(H_M^*,~H_E^*)
.
\]
Now $\tilde{I}(X;Y|Z)$ is of the form 
$\log\det(\id_{n_M+n_E}+C^{-1}D)$, plus some
constant term, where $D \succeq{\bf 0}$. It is known that $\log\det(C)$ is 
concave in $C$ \cite[p.74]{boyd}), thus $\log\det(C^{-1})=-\log\det(C)$ is
convex in $C$, which implies that $\log\det(\id+C^{-1}D)$ is convex. 
Furthermore, it is convex in any block of $C$, thus convex in $A$.
Finally, the set of $A$ such that $\id-AA^*\succ {\bf 0}$ is convex.
\item
{\bf Concavity in $K_X$.}
Recall from (\ref{eq:itilde3}) that 
\begin{eqnarray*}
&&\tilde{I}(X;Y|Z)\\
&=&
\log\det(\id +((H_M^*-H_E^*A^*)(\id-AA^*)^{-1}(H_M-AH_E)+H_E^*H_E)K_X)\\
&&-\log\det(\id+H_EK_XH_E^*).
\end{eqnarray*}
Set
\[
B:=(H_M^*-H_E^*A^*)(\id-AA^*)^{-1}(H_M-AH_E)+H_E^*H_E.
\]
We now have that $\tilde{I}(X;Y|Z)$  is given by
\begin{equation}\label{eq:anotherItilde}
\log\det(\id_n+BK_X)-\log\det(\id_n+H_E^*H_EK_X),
\end{equation}
with $B \succeq H_E^*H_E$. 

If we compute the gradient of (\ref{eq:anotherItilde}) with respect to 
$K_X$, we get that 
\begin{equation}\label{eq:Bg}
(B^{-1}+K_X)^{-1}-((H_E^*H_E)^{-1}+K_X)^{-1} \succeq {\bf 0},
\end{equation}
since $B \succeq H_E^*H_E$.
Recall that 
\[
\frac{\partial(X^{-1})_{kl}}{\partial X_{ij}}=-(X^{-1})_{ki}(X^{-1})_{jl},
\]
so that the derivative of $F:=((H_E^*H_E)^{-1} +K_X)^{-1} $ is a
$n^2\times n^2$ matrix given by
\[
\begin{array}{c}
\left(
\begin{array}{cccc}
-FF_{11}&-FF_{12} & \ldots &-FF_{1n} \\
-FF_{21}&-FF_{22} & \ldots &-FF_{2n} \\
  \vdots&         &        & \vdots  \\
-FF_{n1}&-FF_{n2} &        &-FF_{nn}
\end{array}
\right)
\\
=-((H_E^*H_E)^{-1} +K_X)^{-1}\otimes ((H_E^*H_E)^{-1} +K_X)^{-1}.
\end{array}
\]
To check the concavity in $K_X$, we are thus left to check that 
\[
((H_E^*H_E)^{-1} +K_X)^{-1}\otimes ((H_E^*H_E)^{-1} +K_X)^{-1}\preceq 
(B^{-1}+K_X)^{-1}\otimes (B^{-1}+K_X)^{-1},
\]
which is true by (\ref{eq:Bg}).
\item
Since we have shown above that $\tilde{I}(X;Y|Z)$ is concave in $K_X$ and 
convex in $A$, we have that
\[
\min_{A} \max_{K_X} \tilde{I}(X;Y|Z)= \max_{K_X}\min_{A} \tilde{I}(X;Y|Z).
\]
\end{enumerate}
\end{proof}

From the previous steps of the proof, we now know that
\[
I(X;Y|Z)\leq \max_{K_X} \min_{A} \tilde{I}(X;Y|Z).
\]
We next compute the minimization over $A$. 
Note that we can write $\tilde{I}(X;Y|Z)$ in an alternative way.
Recall that
\[
\tilde{I}(X;Y,Z)=
\log\det(K_{YZ})-\log\det(K_Z)-\log\det(K_{ME}).
\]
By simplifying the Schur complement of $\det(K_{YZ})$ with 
$\det(K_Z)=\det(K_X+K_E)$, we get that 
$\tilde{I}(X;Y|Z)$ is given by
\begin{eqnarray}
&&\log\det(H_MK_XH_M^*+\id_{n_M}-(H_MK_XH_E^*+A)(H_EK_XH_E^*
+\id)^{-1}(H_EK_XH_M^*+A^*))\nonumber\\
&&-\log\det(\id_{n_M}-AA^*).\label{eq:itilde2}
\end{eqnarray}

\begin{prop}
Let $\tilde{A}^*$ be a local minima of $\tilde{I}(X;Y|Z)$.
Then
\[
\tilde{A}^*=(H_E(H_M^*H_M)^{-1}H_M^*V, H_E(H_E^*H_E)^{-1}H_M^*W)(V, W)^{-1},
\]
where $W$ is an arbitrary $n_M\times m$ matrix, $0\leq m \leq n_M$, and 
$V$ is an $n_M\times (n_M-m)$ matrix, such that 
\[
\left(
\begin{array}{c}
V \\
H_E(H_M^*H_M)^{-1}H_M^* V
\end{array}
\right)
\]  
is an invariant subspace of the matrix $M$, as defined in (\ref{eq:M}).
In particular, if $m=n_M$, then
$\tilde{A}^*=H_E(H_E^*H_E)^{-1}H_M^*$. Similarly, if 
$m=0$, then $\tilde{A}^*=H_E(H_M^*H_M)^{-1}H_M^*$.
\end{prop}

\begin{proof}
Let $M_1,M_2,M_3,X$ be square complex matrices. Set 
\[
f(X)=M_1-(X+M_2)M_3(X^*+M_2^*).
\]
It can be shown that 
\[
\nabla_X \log\det(f(X))=-f(X)^{-1}(X+M_2)M_3.
\]
Using this formula, we compute that
\[
\nabla_{A^*}\tilde{I}(X;Y|Z)=0 \iff
f(A)(A^*+H_EK_XH_M^*)^{-1}(H_EK_XH_E^*+\id)=
(\id-AA^*)(A^*)^{-1},
\]
where
\[
f(A)=H_MK_XH_M^*+\id-(H_MK_XH_E^*+A)(H_EK_XH_E^*+\id)^{-1}
(H_EK_XH_M^*+A^*).
\]
This yields the following nonsymmetric algebraic Ricatti equation
$$A^*(H_MK_XH_M^*+\id)^{-1}H_MK_XH_E^*A^*+A^*[(H_MK_XH_M^*+\id)^{-1}]$$
$$+[-H_EK_XH_E^*-\id+H_EK_XH_M^*(H_MK_XH_M^*+\id)^{-1}H_MK_XH_E^*]A^*$$
$$+H_EK_XH_M^*(H_MK_XH_M^*+\id)^{-1}=0. $$
One way of solving an algebraic Riccati \cite{freiling} of the form
\[
{\bf 0}=M_{21}+M_{22}A^*-A^*M_{11}-A^*M_{12}A^*,
\]
is to look for invariant subspaces of
\[
M=
\left(
\begin{array}{cc}
M_{11} & M_{12} \\
M_{21} & M_{22}
\end{array}
\right).
\]
Here we have that $M$ is given by
\begin{equation}\label{eq:M}
\left(\!\!\!\!\!
\begin{array}{cc}
-(H_MK_XH_M^*+\id)^{-1} &-(H_MK_XH_M^*+\id)^{-1}H_MK_XH_E^*\\
H_EK_XH_M^*(H_MK_XH_M^*+\id)^{-1} & -H_EK_XH_E^*-\id+H_EK_XH_M^*(H_MK_XH_M^*+\id)^{-1}H_MK_XH_E^*
\end{array}
\!\!\!\!\!\right).
\end{equation}
Set
\[
F=
\left(
\begin{array}{cc}
H_MK_XH_M^*+\id_{n_M} & 0\\
0 & \id_{n_E}
\end{array}
\right).
\]
We have that $F(M+\id_{n_M+n_E})$ is given by
\[
\left(\!\!\!
\begin{array}{cc}
H_MK_XH_M^* & -H_MK_XH_E^* \\
H_EK_XH_M^*(H_MK_XH_M^*+\id)^{-1} &
-H_EK_XH_E^*+H_EK_XH_M^*(H_MK_XH_M^*+\id)^{-1}H_MK_XH_E^*
\end{array}
\!\!\!\right).
\]
It is easy to see that
\[
F(M+\id)=
\left(
\begin{array}{c}
-H_M \\
-H_E+H_EK_XH_M^*(H_MK_XH_M^*+\id)^{-1}H_M
\end{array}
\right)
(-K_XH_M^*, K_XH_E^*)
\]
which implies that $-1$ is an eigenvalue of $M$. Thus a first
invariant subspace is given by the eigenspace associated to $-1$,
which is the kernel of $M+\id$, or in other words, the subspace 
orthogonal to $(-K_XH_M^*, K_XH_E^*)$:
\[
Ker(M+\id)=
\left(
\begin{array}{c}
U_1\\
H_E(H_E^*H_E)^{-1}H_M^*U_1
\end{array}
\right),
\]
for any $U_1$.
Let us now look for the second invariant subspace.
We first rewrite $M$ as
\[
M=F^{-1}
\left(
\begin{array}{c}
-H_M \\
-H_E+H_EK_XH_M^*(H_MK_XH_M^*+\id)^{-1}H_M
\end{array}
\right)
(-K_XH_M^*, K_XH_E^*)
-\id.
\]
We now show that
\[
\left(
\begin{array}{c}
U_2\\
H_E(H_M^*H_M)^{-1}H_M^*U_2
\end{array}
\right),
\]
is an invariant subspace for any $U_2$.
Indeed, we have that
\[
\begin{array}{l}
F^{-1}
\left(
\begin{array}{c}
-H_M \\
-H_E+H_EK_XH_M^*(H_MK_XH_M^*+\id)^{-1}H_M
\end{array}
\right)\\
=
\left(
\begin{array}{c}
-(H_MK_XH_M^*+\id)^{-1}H_M \\
-H_E+H_EK_XH_M^*(H_MK_XH_M^*+\id)^{-1}H_M
\end{array}
\right)\\
=-
\left(
\begin{array}{c}
\id \\
H_E(H_M^*H_M)^{-1}H_M^*
\end{array}
\right)
(H_MK_XH_M^*+\id)^{-1}H_M
\end{array}
\]
since
\begin{eqnarray*}
&&-H_E(\id-K_XH_M^*(H_MK_XH_M^*+\id)^{-1}H_M)\\
&=&-H_E((H_M^*H_M)^{-1}H_M^*(H_MK_XH_M^*+\id)-K_XH_M^*)
(H_MK_XH_M^*+\id)^{-1}H_M\\
&=&-H_E(H_M^*H_M)^{-1}H_M^*(H_MK_XH_M^*+\id)^{-1}H_M.
\end{eqnarray*}

Thus, a Jordan basis of $M$ is given by
\[
\left(
\begin{array}{cc}
\id_{n_M} & \id_{n_E} \\
H_E(H_M^*H_M)^{-1}H_M^* & H_E(H_E^*H_E)^{-1}H_M^*
\end{array}
\right).
\]
Finally, solutions of the Ricatti equation are given by \cite{freiling} 
\[
\tilde{A}^*=(H_E(H_M^*H_M)^{-1}H_M^*V,H_E(H_E^*H_E)^{-1}H_M^* W)(V, W)^{-1}, 
\]
where $W$ is an $n_M\times m$ matrix, $0\leq m \leq n_M$, and 
$V$ is a $n_M\times n_M-m$ matrix, such that 
\[
\left(
\begin{array}{c}
V  \\
H_E(H_M^*H_M)^{-1}H_M^* V 
\end{array}
\right)
\]
is an invariant subspace of $M$. Note that $W$ can be chosen arbitrary
since $(\id,H_E(H_E^*H_E)^{-1}H_M^*)$ is the eigenspace associated to $-1$.
\end{proof}

\begin{prop}\label{prop:lowrankcon}
Let $\tilde{K}_X$ be an optimal solution to the optimization problem
\begin{eqnarray*}
\max K_X & \min_A \tilde{I}(X;Y|Z) \\
\mbox{s.t.} &  K_X \succeq {\bf 0},~\Tr(K_X)=P,  
\end{eqnarray*}
where $\tilde{A}^*=(H_E(H_M^*H_M)^{-1}H_M^*V,H_E(H_E^*H_E)^{-1} W)(V,W)^{-1}$ 
is the optimal solution for the minimization over $A$.
Then $\tilde{K}_X$ is a low rank matrix.
\end{prop}
\begin{proof}
We have seen in (\ref{eq:anotherItilde}) that $\tilde{I}(X;Y|Z)$ 
can be written
\[
\log\det(\id+BK_X)-\log\det(\id+H_EK_XH_E^*),
\]
where
\[
B:=(H_M^*-H_E^*A^*)(\id-AA^*)^{-1}(H_M-AH_E)+H_E^*H_E.
\]
Using the matrix inversion lemma, we have that
\[
\begin{array}{c}
B^{-1}=(H_E^*H_E)^{-1}-(H_E^*H_E)^{-1}(H_M^*-H_E^*A^*)\cdot\\ 
(\id-AA^*+(H_M-AH_E)(H_E^*H_E)^{-1}(H_M^*-H_E^*A^*))^{-1}
(H_M-AH_E)(H_E^*H_E)^{-1},
\end{array}
\]
so that
\[
\begin{array}{c}
B^{-1}-(H_E^*H_E)^{-1}=-(H_E^*H_E)^{-1}(H_M^*-H_E^*A^*)\cdot\\ 
(\id-AA^*+(H_M-AH_E)(H_E^*H_E)^{-1}(H_M^*-H_E^*A^*))^{-1}
(H_M-AH_E)(H_E^*H_E)^{-1}.
\end{array}
\]
Now
\begin{eqnarray*}
&&(H_E^*H_E)^{-1}(H_M^*-H_E^*A^*) \\      
&=&(H_E^*H_E)^{-1}[H_M^*-H_E^*
   (H_E(H_M^*H_M)^{-1}H_M^*V,H_E(H_E^*H_E)^{-1} W)(V,W)^{-1}]   \\
&=&[(H_E^*H_E)^{-1}H_M^*(V,W)-
   ((H_M^*H_M)^{-1}H_M^*V,(H_E^*H_E)^{-1} W)](V, W)^{-1}\\
&=&(((H_E^*H_E)^{-1}-(H_M^*H_M)^{-1})H_M^*V,{\bf 0})(V, W)^{-1}
\end{eqnarray*}
thus $(H_E^*H_E)^{-1}(H_M^*-H_E^*A^*)$ is low rank and consequently 
$B^{-1}-(H_E^*H_E)^{-1}$ is. 

Now, from Proposition \ref{prop:lowrankach}, we know that either 
$B^{-1}\prec (H_E^*H_E)^{-1} $ and 
$\lambda >0$, or $B^{-1} \succ (H_E^*H_E)^{-1} $ and $\lambda <0$.
This gives a contradiction since $B^{-1}\preceq (H_E^*H_E)^{-1} $, 
implying that $\tilde{K}_X$ has to be low rank. 
\end{proof}
\begin{prop}
Knowing that the rank of $\tilde{K}_X$ is $r<n$, the optimal solution to 
\[
\min_{A} \tilde{I}(X;Y|Z)
\]
is given by
\[
A^*=(H_E(H_M^*H_M)^{-1}H_M^*BH_M U_XV,
H_E(H_E^*H_E)^{-1}H_M^*W)
(BH_M U_XV, W)^{-1}
\]
where $K_X=U_XU_X^*$ and $B=(H_MK_XH_M^*+\id)^{-1}$.
\end{prop}
\begin{proof}
The Jordan decomposition of $M$ is now given by
\[
\begin{array}{c}
M
\left(
\begin{array}{cc}
\id & \id \\
H_E(H_M^*H_M)^{-1}H_M^* & H_E(H_E^*H_E)^{-1}H_M^*
\end{array}
\right)
=\\
\left(
\begin{array}{cc}
\id & \id \\
H_E(H_M^*H_M)^{-1}H_M^* & H_E(H_E^*H_E)^{-1}H_M^*
\end{array}
\right)
\left(
\begin{array}{cc}
J & {\bf 0} \\
{\bf 0} & -\id
\end{array} 
\right).
\end{array}
\]
where
\[
J=-(H_MK_XH_M^*+\id)^{-1}(H_MK_XH_E^*H_E(H_M^*H_M)^{-1}H_M^*+\id).
\]
Let us now look more carefully at $J$.
We first show that when $K_X$ is low rank, $-1$ is an
eigenvalue. Indeed, we have 
\begin{eqnarray*}
&&-(H_MK_XH_M^*+\id)^{-1}(H_MK_XH_E^*H_E(H_M^*H_M)^{-1}H_M^*+\id) +\id \\
&=& -(H_MK_XH_M^*+\id)^{-1}H_MK_X(H_E^*H_E(H_M^*H_M)^{-1}-\id)H_M^*.
\end{eqnarray*}
This is
enough to show that $-1$ is an eigenvalue since $\det(K_X)=0$ by
assumption that $K_X$ is low rank. The above computation also tells us that
\begin{eqnarray*}
&&-(H_MK_XH_M^*+\id)^{-1}(H_MK_XH_E^*H_E(H_M^*H_M)^{-1}H_M^*+\id) \\
&=& -(H_MK_XH_M^*+\id)^{-1}H_MK_X(H_E^*H_E(H_M^*H_M)^{-1}-\id)H_M^*  -\id.
\end{eqnarray*}
Since $K_X$ is low rank, it can be factorized as $K_X=U_XU_X^*$ where 
$U_X$ is a $n\times r$ matrix, if $r<n$ denotes the rank of $K_X$. 
Clearly, $(H_MK_XH_M^*+\id)^{-1}H_MU_X$ is an invariant subspace of 
$J$. A Jordan basis is thus given by 
\[
P=
\left(
\begin{array}{cc}
(H_MK_XH_M^*+\id)^{-1}H_MU_X & Q
\end{array}
\right)
\]
where $Q$ is the eigenspace associated to $-1$. 
Set $B:=(H_MK_XH_M^*+\id)^{-1}$.
This thus gives us a
more precise Jordan basis for $M$ (as defined in (\ref{eq:M})), namely
\[
\begin{array}{c}
\left(
\begin{array}{cc}
P & \id \\
H_E(H_M^*H_M)^{-1}H_M^*P & H_E(H_E^*H_E)^{-1}H_M^*
\end{array}
\right)
=\\
\left(
\begin{array}{ccc}
BH_MU_X & Q & \id \\
H_E(H_M^*H_M)^{-1}H_M^*BH_MU_X
&H_E(H_M^*H_M)^{-1}H_M^* Q
& H_E(H_E^*H_E)^{-1}H_M^*
\end{array}
\right).
\end{array}
\]
In this decomposition, the third block is the eigenspace of $-1$ of
dimension $n_M$ which is always present. The middle block also
corresponds to an eigenspace of $-1$, of dimension $n_M-r$, this one 
appearing only when $K_X$ drops rank. The first block is an invariant subspace,
corresponding to the $r$ eigenvalues of $M$ that are different from $-1$.

From this Jordan basis of $M$, we have that 
\[
A^*=(H_E(H_M^*H_M)^{-1}H_M^*BH_M U_XV,
H_E(H_E^*H_E)^{-1}H_M^*W)
(BH_M U_XV, W)^{-1}
\]
is a solution of the Ricatti equation,  
where $W$ is any $n_M\times (n_M-r)$ matrix, and $V$ is any 
$r\times r$ matrix.
\end{proof}

\subsection{The converse matches the achievability}\label{sec:step3}

So far, we have solved the optimization problem
\[
\min_A\max_{K_X}\tilde{I}(X;Y|Z)
\]
by computing the optimal $\tilde{A}$ in a closed form expression, and
by showing that the optimal $\tilde{K_X}$ is low rank. We are now
ready to conclude the proof, by proving that the optimal $A$ makes the
converse match the achievability.
\begin{prop}
Set $B=(H_MK_XH_M^*+\id)^{-1}$ and let
\[
A^*=(H_E(H_M^*H_M)^{-1}H_M^*BH_M U_XV,
H_E(H_E^*H_E)^{-1}H_M^*W)
(BH_M U_XV, W)^{-1}
\]
be a solution of the Ricatti equation. Then
\[
\tilde{I}(X;Y|Z)=\log \det(\id +H_MK_XH_M^*)-\log \det (\id+H_EK_XH_E^*).
\]
Furthermore, there exists $V,W$ such that $\id-AA^*\succ {\bf 0}$.
\end{prop}
\begin{proof}
Recall from (\ref{eq:itilde1}) that a way of writing $\tilde{I}(X;Y|Z)$ is 
\[
\log\det\left(
\id+
(H_M^*,~H_E^*)
\left(
\begin{array}{cc}
\id &  A \\
A^*   &  \id
\end{array}
\right)^{-1}
\left(
\begin{array}{c}
H_M \\
H_E
\end{array}
\right)
K_X 
\right)
-\log\det(\id+H_EK_XH_E^*),
\]
where
\[
\left(
\begin{array}{cc}
\id &  A \\
A^*   & \id
\end{array}
\right)
=
\left(
\begin{array}{cc}
\id & {\bf 0 } \\
A^* & \id
\end{array}
\right)
\left(
\begin{array}{cc}
\id & {\bf 0 } \\
{\bf 0} & \id-A^*A
\end{array}
\right)
\left(
\begin{array}{cc}
\id & A \\
{\bf 0} & \id
\end{array}
\right).
\]
Thus
\[
\left(
\begin{array}{cc}
\id &  A \\
A^*   &  \id
\end{array}
\right)^{-1}
=
\left(
\begin{array}{cc}
\id & -A \\
{\bf 0} & \id
\end{array}
\right)
\left(
\begin{array}{cc}
\id & {\bf 0 } \\
{\bf 0} & (\id-A^*A)^{-1}
\end{array}
\right)
\left(
\begin{array}{cc}
\id &{\bf 0}  \\
-A^* & \id
\end{array}
\right)
\]
so that
\[
(H_M^*,H_E^*)
\left(
\begin{array}{cc}
\id &  A \\
A^*   &  \id
\end{array}
\right)^{-1}
\left(
\begin{array}{c}
H_M\\
H_E
\end{array}
\right)
=H_M^*H_M+(-H_M^*A+H_E^*)(\id-A^*A)^{-1}(-A^*H_M+H_E)
\]
and
\[
\begin{array}{c}
\tilde{I}(X;Y|Z)=\log\det(\id+H_M^*H_M K_X+
(-H_M^*A+H_E^*)(\id-A^*A)^{-1}(-A^*H_M+H_E) K_X)\\
-\log\det(\id+H_EK_XH_E^*).
\end{array}
\]
We now show that $K_X$ is in the kernel of $-A^*H_M+H_E$.
We have that
\begin{eqnarray*}
(BH_M U_XV, W)^{-1}H_MK_X 
&=&(H_M U_XV, B^{-1}W)^{-1}B^{-1}H_MK_X\\
&=&(H_MU_XV,B^{-1}W)^{-1}H_MU_XU_X^*(H_M^*H_MK_X+\id)\\
&=&
\left(
\begin{array}{c}
V^{-1}U_X^*(H_M^*H_MK_X+\id)\\
{\bf 0}
\end{array}
\right),
\end{eqnarray*}
so that
\begin{eqnarray*}
A^*H_MK_X&=&H_E(H_M^*H_M)^{-1}H_M^*BH_M U_XU_X^*(H_M^*H_MK_X+\id) \\
         &=&H_E(H_M^*H_M)^{-1}H_M^*BB^{-1}H_MK_X\\
         &=&H_EK_X,
\end{eqnarray*}
and thus $A^*H_MK_X=H_EK_X$, so that we get 
\[
\tilde{I}(X;Y|Z)=\log \det(\id +H_MK_XH_M^*)-\log \det (\id+H_EK_XH_E^*).
\]
We now have that
\[
\begin{array}{l}
\id-AA^*\succ {\bf 0} \\
\iff \\
\left(\!\!
\begin{array}{c}
V^*U_X^*H_M^*B^*\\
W^*
\end{array}
\!\!\right)
(BH_MU_XV,W)
-\\
\left(\!\!
\begin{array}{c}
V^*U_X^*H_M^*B^*H_M(H_M^*H_M)^{-1}H_E^* \\ 
W^*H_M(H_E^*H_E)^{-1}H_E^*
\end{array}
\!\!\right)
(H_E(H_M^*H_M)^{-1}H_M^*BH_M U_XV,
H_E(H_E^*H_E)^{-1}H_M^*W)
\succeq {\bf 0}\\
\iff \\
\left(\!\!
\begin{array}{cc}
V^*U_X^*H_M^*B
(\id-H_M(H_M^*H_M)^{-1}H_E^*H_E(H_M^*H_M)^{-1}H_M^*)
BH_MU_XV & {\bf 0} \\
 {\bf 0} &\!\!\!\!\!\!\!\!\!\!\!\!\!\!\!\!\!\!\!\!\!\!\!\! 
W^*(\id-H_M(H_E^*H_E)^{-1}H_M^*)W
\end{array}
\!\!\right)\succ {\bf 0},
\end{array}
\]
since 
\begin{eqnarray*}
&&V^*U_X^*H_M^*B^*W-V^*U_X^*H_M^*B^*H_M(H_M^*H_M)^{-1}H_M^*W\\
&=&V^*U_X^*[H_M^*B^*-H_M^*B^*H_M(H_M^*H_M)^{-1}H_M^*]W={\bf 0}.
\end{eqnarray*}
To conclude the proof, notice that we have
\[
\id-H_M(H_M^*H_M)^{-1}H_E^*H_E(H_M^*H_M)^{-1}H_M^* \preceq{\bf 0} 
\iff H_M^*H_M \preceq H_E^*H_E
\]
and
\[
\id-H_M(H_E^*H_E)^{-1}H_M^*\preceq {\bf 0}
\iff H_E^*H_E \prec H_M^*H_M.
\]
Thus if $H_M^*H_M-H_E^*H_E$ is indefinite,
there exists $V$ and $W$ such that the above matrix is positive 
definite.
\end{proof}

%
%

\section{Conclusion}

In this paper, we considered the problem of computing the perfect secrecy
capacity of a multiple antenna channel, based on a generalization of
the wire-tap channel to a MIMO broadcast wire-tap channel. We proved
that for an arbitrary number of transmit/receive antennas, the perfect
secrecy capacity is the difference of the two capacities, the one of
the legitimate user minus the one of the eavesdropper.

%
%

\section*{Appendix}

\begin{prop}\label{prop:maxgauss}
Let $A$,$B$ be circularly symmetric complex jointly Gaussian random 
vectors with strictly positive definite 
covariance matrices. Let $X$ be a random vector independent of $A$ and 
$B$, and $S$ be a positive definite matrix. The optimal solution to 
\begin{eqnarray*}
\max \Pc(X) & h(X+A,X+B)-h(X+B)\\
\mbox{s.t.} & \Tr(K_X)=P
\end{eqnarray*}
is Gaussian.
\end{prop}
\begin{proof}
First note that
\[
\frac{1}{\sqrt{2}}
\left(
\begin{array}{cc}
\id & -\id \\
\id & \id
\end{array}
\right)
\left(
\begin{array}{c}
X+A \\
X+B
\end{array}
\right)
=
\left(
\begin{array}{c}
\frac{1}{\sqrt{2}}(A-B)\\
\sqrt{2}X+\frac{1}{\sqrt{2}}(A+B)
\end{array}
\right).
\]
Since multiplication by a unitary matrix does not change the entropy,
\begin{eqnarray*}
&&h(X+A,X+B) \\
&=&
h\left(\sqrt{2}X+\frac{1}{\sqrt{2}}(A+B),\frac{1}{\sqrt{2}}(A-B)\right)\\
& =& h\left(\sqrt{2}X+\frac{1}{\sqrt{2}}(A+B)|\frac{1}{\sqrt{2}}(A-B)\right)\\
&&   +h\left(\frac{1}{\sqrt{2}}(A-B)\right)\\
&=& h(\sqrt{2}X+U)+h\left(\frac{1}{\sqrt{2}}(A-B)\right)
\end{eqnarray*}
where $U$ is Gaussian with covariance matrix $K_U$ given by
\[
\begin{array}{c}
\frac{1}{2}E[(A+B)(A+B)^*]-\frac{1}{2}E[(A+B)(A-B)^*]\cdot  \\
E[(A-B)(A-B)^*]^{-1}E[(A-B)(A+B)^*],
\end{array}
\]
using conditional Gaussian distribution. 

To maximize
\[
h(X+A,X+B)-h(X+B),
\]
we thus need to maximize
\[
 h(\sqrt{2}X+U)-h(X+B),
\]
or equivalently
\[
 h(X+U')-h(X+B)
\]
where $U'=U/\sqrt{2}$ is Gaussian, independent of $X$.
The optimal distribution of such expression has been shown 
to be Gaussian by Liu and Viswanath \cite{Liu} in the case of real
Gaussian vectors. Their result can be readily extended to 
the circularly symmetric complex Gaussian case.
\end{proof}

\begin{lemma}\label{lem:spd}
If $A=A^* \succ {\bf 0}$ and $B=B^* \succ {\bf 0}$, then the matrix 
$AB$ has all positive eigenvalues.
\end{lemma}
\begin{proof}
Since $A\succ {\bf 0}$, we can write $A = A^{1/2}(A^*)^{1/2}$ with 
$A^{1/2}$ invertible. Therefore, 
\[
AB = A^{1/2}( (A^*)^{1/2} B A^{1/2} ) A^{-1/2},
\]
has the same eigenvalues as the matrix $(A^*)^{1/2} B A^{1/2}$, which
is positive definite.
\end{proof}

%
%


\begin{thebibliography}{99}
%
\bibitem{barros}
J. Barros and M. R. D. Rodrigues, ``Secrecy Capacity of Wireless Channels'', 
{\em IEEE International Symposium on Information Theory}, Seattle, July 2006.
%
\bibitem{bloch}
 M. Bloch, J. Barros, M. R. D. Rodrigues, S. W. McLaughlin, 
``Wireless Information-Theoretic Security - Part I: Theoretical Aspects''. 
Submitted to {\em IEEE Transactions on Information Theory}, Special 
Issue on Information-Theoretic Security, November 2006
%
\bibitem{boyd}
S. Boyd and L. Vandenberghe, ``Convex Optimization'', 
Cambridge University Press, 2004.
%
\bibitem{freiling}
G. Freiling, ``A Survey on Nonsymmetric Ricatti Equations'', 
{\em Lin. Algebra and its Appl.}, 251-252, 2002.
%
\bibitem{elgamal}
P. Gopala, L. Lai, and H. El Gamal, ``On the Secrecy Capacity of
Fading Channels'', submitted to {\em IEEE Transactions on Information 
Theory}, Oct. 2006
%
\bibitem{Khisti}
A. Khisti, G. Wornell, A. Wiesel, Y. Eldar, ``On the Gaussian MIMO 
Wiretap Channel'', in {\em Proc. of IEEE International Symposium on
  Information Theory}, Nice, 2007.
%
\bibitem{Hellman}
S.K. Leung-Yan-Cheong, M.E. Hellman, ``The Gaussian Wire-Tap
Channel'', {\em IEEE Trans. on Information Theory}, vol. 24, July 1978.
%
\bibitem{hero}
A. O. Hero, ``Secure Space-Time Communication," , {\em IEEE Trans. on 
Info Theory}, Vol. 49, No. 12, pp. 1-16, Dec. 2003.
%
\bibitem{Parada}
P. Parada, R. Blahut,``Secrecy capacity of SIMO and slow fading
channels,'' in {\em Proc. of IEEE International Symposium on
  Information Theory}, Adelaide, 2005.
%
\bibitem{Li2}
Z. Li, R. Yates, W. Trappe,``Secrecy capacity of independent parallel
channels'', in {\em Proc. of Allerton conference}, 2006. 
%
\bibitem{Li}
Z. Li, W. Trappe, R. Yates, ``Secret communication via multi-antenna 
transmission'', in the proceedings of {\em Conference on Information 
Sciences and Systems (CISS)}, March 2007.
%
\bibitem{Poor}
Y. Liang, H. V. Poor, ``Secure Communication over Fading Channels'', 
in {\em Proc. of Allerton}, 2006.
%
\bibitem{liang}
Y. Liang, H. V. Poor, Shlomo Shamai (Shitz),
``Secure Communication over Fading Channels'', 
Submitted to {\em IEEE Transactions on Information Theory}, Special
Issue on Information Theoretic Security, November 2006
%
\bibitem{poormimo}
R. Liu, H. V. Poor, 
``Multiple Antenna Secure Broadcast over Wireless Networks'',
to appear in the Proceedings of the First International Workshop on 
Information Theory for Sensor Networks, Santa Fe, NM, June 2007.
%
\bibitem{Liu}
T. Liu, P. Viswanath, ``An Extremal Inequality Motivated by
Multiterminal Information Theoretic Problems'', to appear in 
{\em IEEE Transactions on Information Theory}.
%
\bibitem{Wyner}
A.D. Wyner, ``The wire-tap channel,'' {\em Bell. Syst. Tech. J.},
vol. 54, October 1975.
%
\end{thebibliography}
\end{document}